\documentclass[prd,aps,twocolumn,superscriptaddress,showpacs,dvips]{revtex4}
\usepackage{feynmp}
\usepackage{amssymb}
\usepackage{epsfig}
\usepackage{graphicx}
\usepackage{subfigure}
\usepackage{hyperref}

\begin{document}

\title{Unconventional behavior of Dirac fermions in three-dimensional gauge theory}

\author{Jing Wang}
\address{\small {\it Department of Modern Physics, University of
Science and Technology of China, Hefei, Anhui, 230026, P.R. China}}
\author{Guo-Zhu Liu}
\address{\small {\it Max Planck Institut f$\ddot{u}$r Physik komplexer
Systeme, D-01187 Dresden, Germany}} \address{\small {\it Department
of Modern Physics, University of Science and Technology of China,
Hefei, Anhui, 230026, P.R. China}}

\begin{abstract}
We study unconventional behavior of massless Dirac fermions due to
interaction with a U(1) gauge field in three time-space dimensions.
At zero chemical potential, the longitudinal and transverse
components of gauge interaction are both long-ranged. There is no
fermion velocity renormalization since the system respects Lorentz
invariance. At finite chemical potential, the Lorentz invariance is
explicitly broken by the finite Fermi surface. The longitudinal
gauge interaction is statically screened and becomes unimportant,
whereas the transverse gauge interaction remains long-ranged and
leads to singular renormalization of fermion velocity. The anomalous
dimension of fermion velocity is calculated by means of the
renormalization group method. We then examine the influence of
singular velocity renormalization on several physical quantities,
and show that they exhibit different behavior at zero and finite
chemical potential.
\end{abstract}

\pacs{11.10.Hi, 11.10.Kk, 71.10.Hf}

\maketitle


\section{Introduction}

Quantum electrodynamics of massless Dirac fermions defined in three
time-space dimensions (QED$_3$) has been extensively investigated
for 30 years \cite{Pisarski, Appelquist86, Appelquist88, Nash,
Dagotto, Atkinson, Pennington, Roberts, Maris95, Appelquist95,
Gusynin96, Maris96, Hands02, Liu03, Fischer, Bashir, Feng1, Feng2,
Bashir08, Liu10, WangLiu1, WangLiu2, WangLiu3}. Different from its
four-dimensional counterpart, QED$_3$ is a super-renormalizable
field theory, so its ultraviolet behavior can be well controlled.
This gauge field theory is known to exhibit asymptotic freedom
\cite{Appelquist95}, which means the gauge interaction becomes
stronger at lower energies. Because of this feature, intriguing
non-perturbative phenomena are expected to occur in the low-energy
regime. When Dirac fermions are strictly massless, the Lagrangian
respects a continuous chiral symmetry. However, the strong gauge
interaction may trigger fermion-antifermion vacuum condensation,
$\langle \bar{\psi}\psi\rangle \neq 0$, which generates a finite
fermion mass and induces dynamical chiral symmetry breaking (DCSB).
Appelquist \emph{el al.} first found that DCSB can occur when the
fermion flavor $N$ is smaller than some critical value $N_c =
32/\pi^2$ \cite{Appelquist88}. Motivated by this very interesting
prediction, intense theoretical effort has been devoted to studying
this problem \cite{Nash, Dagotto, Atkinson, Pennington, Maris95,
Appelquist95, Gusynin96, Maris96, Hands02, Liu03, Fischer, Bashir,
Feng1, Feng2, Bashir08, Liu10}. Despite some debate \cite{Pisarski,
Atkinson, Pennington}, most of these analytical and numerical
calculations agree that a critical value exists at roughly $N_c
\approx 3.5$. A remarkable consequence of DCSB is that it leads to
weak confinement \cite{Roberts, Maris95}.

Apart from exhibiting many properties interesting in the context of
elementary particle physics, QED$_3$ also has extensive applications
in condensed matter physics. Specifically, it serves as an effective
low-energy theory of $d$-wave high-temperature superconductor
\cite{Lee05, Affleck, Ioffe, Kim97, Kim99, Rantner, Rantner2, Franz,
Franz2, Herbut, Hermele, Kaul, Liu02, Liu03, Liu05, Liu09} and some
quantum spin liquid state \cite{Ran07}. The occurrence of DCSB in
QED$_3$ corresponds to the formation of two-dimensional long-range
antiferromagnetic order \cite{Kim99, Franz2, Herbut, Liu03, Liu05}.
Recently, it has been proposed that QED$_3$ may be simulated on
optical lattice \cite{Kapit}, which provides an opportunity of
measuring DCSB experimentally.

When the fermion flavor is beyond the threshold, $N > N_c$, no
dynamical fermion mass can be generated. There is no fermion vacuum
condensation, i.e., $\langle \bar{\psi}\psi\rangle = 0$, and the
continuous chiral symmetry is preserved. However, the absence of
chiral condensation in the vacuum does not mean that the chiral
symmetric phase is trivial. On the contrary, many highly nontrivial
features can emerge in the symmetric phase of QED$_3$. In
particular, the gauge interaction is able to cause breakdown of
Fermi liquid (FL). In 1973, Holstein \emph{et} \emph{al.} showed
\cite{Holstein} that the unscreened transverse component of
electromagnetic field in (3+1)-dimensional non-relativistic electron
gas leads to unusual logarithmic specific heat, $\propto T\ln T$,
which is apparently out of the scope of FL theory. This discovery
have stimulated extensive investigations of non-FL behavior in
various gauge theories, in the contexts of both condensed matter
physics \cite{Reizer, Lee92, Gan, Khve, Nayak, Polchinski,
Altshuler, Halperin, Ichinose, IchinoseOnoda, Kim97, Varma, Senthil}
and particle physics \cite{Baym, Vega, Schafer, Ipp1, Ipp2}. The
non-FL behavior of massless Dirac fermions in QED$_3$ has also been
discussed \cite{Franz, WangLiu1, WangLiu2}. In addition, QED$_3$ can
be used to describe some intriguing states, such as algebraic spin
liquid state \cite{Rantner, Rantner2, Franz, Franz2, Herbut,
Hermele} and algebraic charge liquid state \cite{Kaul}.

In some many-particle systems described by QED$_3$, there is a
finite density of massless Dirac fermions. The finite fermion
density is usually represented by chemical potential $\mu$. An
interesting question is how this chemical potential affects the
physical properties of QED$_3$. The impacts of finite chemical
potential on DCSB have been addressed in \cite{Feng1, Feng2, Liu10},
where it is found that the critical flavor $N_c$ is lowered as
chemical potential grows and DCSB is completely suppressed when
chemical potential is sufficiently large. In the chiral symmetric
phase with strictly massless Dirac fermions, the gauge interaction
also leads to different properties at zero and finite chemical
potential. For instance, the Dirac fermion damping rate behaves as
$\propto \omega^{1/2}$ at zero chemical potential \cite{WangLiu1}
and $\propto \omega^{2/3}$ at finite chemical potential
\cite{WangLiu2}.

In this paper, we consider chiral symmetric phase of QED$_3$ and
study unconventional behavior of massless Dirac fermions at finite chemical
potential. We assume a relatively large fermion flavor $N$ so that the
fermions are strictly massless. One important effect of a finite chemical
potential is that it explicitly breaks the Lorentz invariance. As a
consequence, the longitudinal component of gauge field develops an
effective mass that is proportional to the chemical potential $\mu$,
which is analogous to the static screening of Coulomb interaction.
However, the transverse component of gauge field remains massless
due to the gauge invariance. Because of such difference in the
longitudinal and transverse components of gauge field, the temporal and
spatial components of fermion self-energy are no longer identical,
which in turn gives rise to nontrivial renormalization of fermion
velocity.

We shall analyze fermion velocity renormalization by performing a
renormalization group (RG) calculation \cite{Shankar94, Son2007,
Huh, WLK}. The fermion velocity remains a constant at zero chemical
potential after including the gauge interaction corrections since
the Lorentz invariance is preserved. However, it acquires strong
momentum dependence after developing an anomalous dimension
$\gamma_v$ at finite chemical potential. The appearance of nonzero
anomalous dimension is a consequence of Lorentz symmetry breaking
and gauge symmetry. Therefore, QED$_3$ with a finite chemical
potential is fundamentally different from that defined at zero
chemical potential. We then evaluate specific heat, density of
states (DOS), and compressibility of massless fermions using
$\gamma_v$ obtained in the RG analysis. After analytical and
numerical calculations, we demonstrate that massless Dirac fermions
exhibits unconventional, non-FL like, behavior at finite chemical
potential.

The rest of this paper is organized as follows. We define the
Lagrangian and then perform RG analysis of fermion velocity
renormalization in Sec.~\ref{sec_anomous_v_F}. A finite anomalous
dimension of fermion velocity is obtained. We calculate specific
heat in Sec.~\ref{sec_specific_heat}, and DOS and compressibility in
Sec.~\ref{sec_DOS_kappa}. In section \ref{sec_summary}, we briefly
summarize the results.

\section{Renormalization group analysis of Fermi velocity renormalization}
\label{sec_anomous_v_F}

The Lagrangian density for $\mathrm{QED_{3}}$ with $N$-flavor Dirac
fermions is given by
\begin{eqnarray}
\mathcal{L} \!= \!\!\sum^N_{i=1}\!\bar{\psi}_i[(\partial_\tau \!-\! \mu \!
- \!iea_0)\gamma_0 \!-\! iv_F\mathbf{\gamma} \cdot (\partial\!-\! ie\mathbf{a}
)]\psi_i\!-\!\frac{1}{4}F^2_{\mu\nu},\label{eq_model_QED3}
\end{eqnarray}
where $\mu$ represents the chemical potential and $v_F$ is the
constant fermion velocity. The Dirac fermions can be described by a
four-component spinor field $\psi$ and
$\bar{\psi}=\psi^\dagger\gamma^0$. The $4 \times 4$ $\gamma$
matrices can be chosen as:
$\gamma_\mu=(\sigma_3,\sigma_1,\sigma_2)\otimes\sigma_3$, which
satisfy the standard Clifford algebra
$\{\gamma_\mu,\gamma_\nu\}=2g_{\mu\nu}$ with metric $g_{\mu\nu} =
\mathrm{diag}(1,1,1)$. In this paper, we consider a large $N$ and
perform $1/N$ expansion. For convenience, we work in units with
$\hbar = k_{B} = 1$ and restore them whenever necessary.

It is now helpful to further remark on the physical meaning of
fermion velocity $v_F$. There are two ways to define the velocity
$v_F$. If we regard QED$_3$ as a standard relativistic quantum field
theory, the velocity of massless particles is simply the velocity of
light, i.e., $v_F \equiv c$. If, on the other hand, we consider the
effective QED$_3$ theory that is derived from a microscopic model of
some condensed matter system (for instance, Hubbard model of
high-temperature superconductor) \cite{Lee05, Affleck, Ioffe}, the
fermion velocity should be calculated from the band structure of the
corresponding microscopic model. In the latter case, the fermion
velocity $v_F$ is no longer equal to the velocity of light, $c$. Its
magnitude is strongly material dependent, but is always much smaller
than $c$ \cite{Lee05, Affleck, Ioffe}. In the present paper, we
assume a constant velocity $v_F$ and study its renormalization due
to gauge interaction at finite chemical potential. The main
conclusion depends only on the momentum dependence of the
renormalized velocity, but not on the concrete magnitude of the
constant $v_F$.

In Euclidian space, the free propagator of massless Dirac fermions
at zero $\mu$ is
\begin{equation}\label{fermion_propagator}
G_0(k) = \frac{1}{k\!\!\!/} =
\frac{\gamma_0k_0+v_{F}\gamma\cdot\mathbf{k}}{k^2}.
\end{equation}
At finite $\mu$, it proves convenient to work in the Matsubara
formalism and write the fermion propagator as
\begin{equation}\label{eq:add_mu}
G_{0}(i\omega_{n},\mathbf{k}) =\frac{1}{(i\omega_{n}+\mu)\gamma_0
-v_{F}\mathbf{\gamma}\cdot \mathbf{k}},
\end{equation}
where the fermion frequency is $i\omega_n = i(2n+1)\pi/\beta =
i(2n+1)\pi T$ with $n$ being integers. For notational convenience,
we use $k_0$ to denote the imaginary frequency $i\omega_n$, so that
the fermion propagator can also be written as
\begin{equation}
G_0(k) = \frac{1}{k\!\!\!/} =
\frac{\gamma_0(k_0+i\mu)+v_{F}\gamma\cdot\mathbf{k}}{k^2}.
\end{equation}

\subsection{$\mu=0$}

We first consider the case of zero $\mu$. The key physical quantity
is the fermion self-energy function, which can be used to calculate
the RG flow of fermion velocity $v_F$ \cite{Son2007, Huh, WLK}.
Following the RG strategy presented in Ref.~\cite{Son2007}, we
introduce two cutoffs $\Lambda_0$ and $\Lambda_1$, with $\Lambda_1$
being smaller than $\Lambda_0$. To the leading order of $1/N$
expansion, the one-loop fermion self-energy at zero $\mu$ is
\begin{eqnarray}\label{self_energy_mu_eq_0}
\Sigma(k) &=& \frac{\alpha}{N}
\int_{\Lambda_1}^{\Lambda_0}\frac{d^3q}
{(2\pi)^3}\gamma_{\mu}\frac{( k\!\!\! / - q\!\! \! /)}{(k-q)^{2}}
\gamma_{\nu}
D_{\mu\nu}(q)\nonumber\\
&\equiv& \gamma_{0}k_{0}\Sigma_{0}+ v_{F}\gamma_{i}k_{i}\Sigma_{1}.
\end{eqnarray}
The full gauge field propagator $D_{\mu\nu}(q)$ is given by the
Dyson equation
\begin{equation}\label{eq:photon_full_0}
D_{\mu\nu}^{-1}(q) = D_{\mu\nu}^{(0)-1}(q)+ \Pi_{\mu\nu}(q),
\end{equation}
with free gauge field propagator
\begin{equation}\label{eq:photon_0}
D_{\mu\nu}^{(0)}(q) = \frac{1}{q^{2}}\left(g_{\mu\nu}
-\frac{q_{\mu}q_{\nu}}{q^2}\right),
\end{equation}
in the Landau gauge. To the leading order of $1/N$ expansion, the
one-loop contribution to vacuum polarization tensor $\Pi_{\mu\nu}$
is
\begin{eqnarray}\label{eq:pi}
\Pi_{\mu\nu}(q) &=& - \alpha\int \frac{d^{3}k}{(2\pi)^{3}}
\frac{\mathrm{Tr}[\gamma_{\mu}k\!\!\!/\gamma_{\nu} (q\!\!\!/ +
k\!\!\!/)]}{k^2(q+k)^2} \nonumber \\
&=& \Pi(q^2)\left(g_{\mu\nu} - \frac{q_{\mu}q_{\nu}}{q^2}\right).
\end{eqnarray}
Here, it is convenient to define $\alpha = Ne^2$, which is fixed as
$N$ taken to be large \cite{Appelquist88}. Now the gauge field
propagator becomes
\begin{equation}\label{photon_propagtor_mu_eq_0}
D_{\mu\nu}(q) = \frac{1}{q^{2}+\Pi(q)}\left(g_{\mu\nu}-
\frac{q_{\mu}q_{\nu}}{q^2}\right),
\end{equation}
where $\Pi(q) = \frac{\alpha q}{8}$ at zero temperature.

According to the Dyson equation, $G^{-1}(k) = G_0^{-1}(k) -
\Sigma(k)$, the full fermion propagator is
\begin{eqnarray}
G(k) &=& \frac{1}{k\!\!\!/ - \Sigma(k)} \nonumber \\
&=& \frac{1}{\left(1-\Sigma_{0}\right)\gamma_{0}k_{0} +
\left(1-\Sigma_{1}\right)v_F \gamma_{i}k_{i}}.
\end{eqnarray}
Apparently, $\left(1-\Sigma_0\right)$ correspond to the wave
function renormalization, whereas $\left(1-\Sigma_1\right)$
represents the product of the renormalization factors of wave
function and fermion velocity.

By inserting Eq.~(\ref{photon_propagtor_mu_eq_0}) into Eq.
(\ref{self_energy_mu_eq_0}), it is easy to find that $\Sigma_{0} =
\Sigma_{1}$. Apparently, the temporal and spatial parts of fermion
propagator are equally renormalized, which originates from the
Lorentz invariance of the system at zero $\mu$. Therefore, the
fermion velocity $v_F$ is a flow-invariant constant.

\subsection{$\mu\neq0$}

At finite chemical potential, QED$_3$ theory can exhibit
qualitatively new features. First of all, there appears a finite
Fermi surface , which explicitly breaks the Lorentz invariance. In
this case, the temporal and spatial components of fermion
self-energy are no longer equivalent, so the fermion velocity may be
singularly renormalized.

To proceed, we first need to discuss the effects of a finite
chemical potential $\mu$ on the effective gauge interaction
function. These effects are reflected in the vacuum polarization
function $\Pi(q,\mu)$. It is technically hard to obtain an entirely
analytical expression for $\Pi(q,\mu)$, so we will derive an
approximate $\Pi(q,\mu)$ that captures the essential features of
QED$_3$ at finite $\mu$.

Under the Coulomb gauge condition, $q_{i}\Pi_{ij}(q) =
0$, the vacuum polarization tensor can be decomposed into two
independent parts \cite{Dorey}:
\begin{equation}
\Pi_{\mu\nu}(q_0,\mathbf{q},\beta) =
\Pi_{A}(q_0,\mathbf{q},\beta) A_{\mu\nu} +
\Pi_{B}(q_0,\mathbf{q},\beta)
B_{\mu\nu},\nonumber \\
\end{equation}
where
\begin{eqnarray}
A_{\mu\nu} &=& \delta_{\mu 0}\delta_{0\nu}, \\
B_{\mu\nu} &=& \delta_{\mu i}\Big(\delta_{ij} -
\frac{q_{i}q_{j}}{\mathbf{q}^{2}}\Big)\delta_{j\nu}, \,\,\,\, \{i,j
= 1,2\}.
\end{eqnarray}
They are orthogonal and related by
\begin{equation}
A_{\mu\nu} + B_{\mu\nu} = \delta_{\mu\nu} -
\frac{q_{\mu}q_{\nu}}{q^{2}}.
\end{equation}
Now, the full gauge field propagator $D_{\mu\nu}(q_0, {\bf
q}, \beta)$ can be written as
\begin{equation}\label{eq:photon_full}
D_{\mu\nu}(q_0, {\bf q}, \beta) =
\frac{A_{\mu\nu}}{\mathbf{q}^{2} + \Pi_{A}(q_0, {\bf q},
\beta)}+ \frac{B_{\mu\nu}}{q^{2} + \Pi_{B}(q_0, {\bf q},
\beta)},
\end{equation}
where the functions $\Pi_{A}$ and $\Pi_{B}$ are related to the
temporal and spatial components of vacuum polarization tensor
$\Pi_{\mu\nu}$ by
\begin{eqnarray}
\Pi_{A} &=& \Pi_{00}, \\
\Pi_{B} &=& \Pi_{ii} - \frac{q_0^{2}}
{\mathbf{q}^{2}}\Pi_{00},
\end{eqnarray}
with $q^{2} = q_0^{2} + {\bf q}^{2}$. Based on these
quantities, it is convenient to write the fermion self-energy as
\begin{eqnarray}\label{definition_sigma}
\Sigma(k) &\equiv&\Sigma_{A}+\Sigma_{B}
\equiv\gamma_{0}k_{0}\Sigma_{0}+ v_{F}\gamma_{i}k_{i}\Sigma_{1},
\end{eqnarray}
where
\begin{eqnarray}\label{definition_sigma_0_1}
\Sigma_{0} &=& \Sigma_{A0} + \Sigma_{B0}, \\
\Sigma_{1} &=& \Sigma_{A1}+\Sigma_{B1}.
\end{eqnarray}

In an earlier publication \cite{WangLiu2}, the full analytical
expressions of polarization function $\Pi(q,\mu)$ was derived. These
analytical expressions are too complicated and can not be directly
used. It is necessary to make proper approximations. We first
consider the temporal component $\Pi_{00}(q,\mu)$. In order to
simplify computations, we utilize the so-called instantaneous
approximation \cite{Dorey}, i.e., $q_0 = 0$, and write
$\Pi_{00}(q,\mu)$ approximately as
\begin{eqnarray}\label{pi00}
\Pi_{00}(\mathbf{q},\mu) &=& \left\{\begin{array}{ll}
\frac{\alpha\mu}{\pi},\hspace{0.5cm} \mu\geq\frac{|\mathbf{q}|}{2}
\\
\frac{\alpha |\mathbf{q}|}{8}, \hspace{0.3cm}\mu<\frac{|\mathbf{q}
|}{2}.
\end{array}\right.
\end{eqnarray}
Apparently, the finite $\mu$ serves as an energy scale:
$\Pi_{00}(q,\mu)$ behaves quite differently above and below $2\mu$.
In the low-energy (long wavelength) limit, $\mathbf{q} \rightarrow
0$, we have
\begin{eqnarray}
\Pi_{00}(\mathbf{q} \rightarrow 0,\mu) \neq 0.
\end{eqnarray}
As a consequence, the temporal part of gauge interaction becomes
short-ranged due to static screening. In other words, the temporal
component of gauge field acquires a finite effective mass that is
proportional to $\mu$. Such short-ranged interaction does not lead
to any singular contribution to fermion self-energy, and thus can be
simply neglected.

We then consider the spatial component of polarization function
$\Pi_{ii}(q,\mu)$. Similar to its temporal counterpart,
$\Pi_{ii}(q,\mu)$ also exhibit different behavior above and below
the energy scale $2\mu$. When $|\mathbf{q}| > 2\mu$, we still use
the instantaneous approximation and have
\begin{eqnarray}
\Pi_{ii}(\mathbf{q},\mu) = \frac{\alpha |\mathbf{q}|}{8}.
\end{eqnarray}
When $|\mathbf{q}| < 2\mu$, $\Pi_{ii}(q,\mu)$ vanishes at zero
energy $q_0 = 0$, so it is not appropriate to use the instantaneous
approximation. The energy dependence of $\Pi_{ii}(q,\mu)$ should be
explicitly maintained \cite{WangLiu2}. As pointed out in
Ref.~\cite{WangLiu2}, the fermion self-energy is dominated by the
low-energy region of $q_0 \ll |\mathbf{q}| \ll 2\mu$. We notice this
approximation is widely used in the calculations of fermion
self-energy due to gauge interaction \cite{Holstein, Lee92, Khve,
Polchinski, Altshuler} and critical ordering fluctuation
\cite{Metzner}. In this region, $\Pi_{ii}(q,\mu)$ can be
approximated as
\begin{eqnarray}\label{eq_Pi_ii}
\Pi_{ii}(q_0,\mathbf{q},\mu) =
\frac{\alpha\mu}{2\pi}\frac{q_0}{|\mathbf{q}|}.
\end{eqnarray}
At the lowest energy $q_0 = 0$, we have
\begin{eqnarray}
\Pi_{ii}(q_0 = 0,\mathbf{q},\mu) = 0.
\end{eqnarray}
This fact implies that, the transverse component of gauge
interaction remains long-ranged even when the dynamical screening
effect is taken into account. It also explains why the instantaneous
approximation can not be used in this region. Physically, the
long-range property of the transverse gauge interaction is protected
by the local gauge invariance.

The fermion self-energy is given by
\begin{eqnarray}
&&\Sigma_{B} \nonumber\\
&=& \frac{\alpha}{N}\int_{\Lambda_1}^{\Lambda_0}
\frac{d^3q}{(2\pi)^3}\gamma_{\mu}\frac{(k\!\!\! /- q\!\! \!
/)}{(k-q)^{2}}\gamma_{\nu}\frac{B_{\mu\nu}}{q^2+\Pi_B} \nonumber\\
&=& \frac{\alpha}{N}\int_{\Lambda_1}^{\Lambda_0}
\frac{d^3q}{(2\pi)^3}\frac{-\gamma_{0}(k_0- q_0)+ v_{F}\gamma \cdot
(\mathbf{k- q})}{(k-q)^{2}} \frac{1}{q^2+\Pi_{ii}} \nonumber
\\&& + \frac{\alpha}{N} \int_{\Lambda_1}^{\Lambda_0}
\frac{d^3q}{(2\pi)^3}\frac{-2v_{F}\gamma_{1}k_1 q_1^2 - 2v_{F}\gamma_{2}
k_2q_2^2}{\mathbf{q}^2(k-q)^{2}}\frac{1}{q^2
+ \Pi_{ii}}\nonumber\\\label{Sigma_B_mu_neq_0} \\
&=& (\gamma_{0}k_{0}\Sigma_{B0}+ v_{F}\gamma_{i}k_{i}
\Sigma_{B1} )_{\mathrm{fir}}\nonumber\\
&&+(\gamma_{0}k_{0}\Sigma_{B0}+ v_{F}\gamma_{i}k_{i}
\Sigma_{B1})_{\mathrm{sec}}.
\label{self_energy_mu_neq_0}
\end{eqnarray}
According to the general RG scheme \cite{Shankar94}, one needs to
integrate out the high-energy degrees of freedom (fast modes) step
by step, until eventually reaching the lowest energy. Since
$\Pi_{ii}$ behaves differently at high and low energies, the
self-energy should be calculated separately.

For small external momentum $k\ll \Lambda_1$, we are allowed to make
the approximation \cite{Son2007}, $(k-q)^{2}\approx q^2$, and write
the first term of Eq. (\ref{self_energy_mu_neq_0}) as
\begin{eqnarray}
(\Sigma_{B1}-\Sigma_{B0})_{\mathrm{fir}}
&\approx&\frac{2\alpha}{N}
\int_{\Lambda_1}^{\Lambda_0}\frac{d^3q}{(2\pi)^3}
\frac{1}{q^{2}}\frac{1}{(q^2+\Pi_{ii})}\nonumber \\
&\approx& \frac{2\alpha}{N}\int^{\Lambda_0}_{\Lambda_1}
\frac{d^3q}{(2\pi)^3}\frac{1}{q^2} \frac{1}{q^2+\frac{\alpha
\mid\mathbf{q}\mid}{8}}.
\end{eqnarray}
From RG theory, we know that the fermion velocity can receive
singular renormalization only when the self-energy contains a
logarithmic term. One can check that there is no such term in the
regime where $\frac{\alpha\mid\mathbf{q}\mid}{8}\ll q^2$.
Nevertheless, a logarithmic term emerges as one goes to the
low-energy regime where $\frac{\alpha\mid\mathbf{q}\mid}{8}\gg q^2$.
After neglecting $q^2$, it is straightforward to obtain
\begin{eqnarray}
(\Sigma_{B1}-\Sigma_{B0})_{\mathrm{fir}} &\approx&
\frac{2\alpha}{N}\int^{\Lambda_0}_{\Lambda_1}
\frac{d^3q}{(2\pi)^3}\frac{1}{q^2} \frac{1}
{\frac{\alpha \mid\mathbf{q}\mid}{8}} \nonumber\\
&=& \frac{4}{\pi N}\ln\frac{\Lambda_0} {\Lambda_1}.\label{first}
\end{eqnarray}
Paralleling the above analysis, the second term of Eq.
(\ref{Sigma_B_mu_neq_0}) can be computed in an analogous manner:
\begin{eqnarray}
&& -2\frac{\alpha}{N}\int\frac{d^3q}{(2\pi)^3}\frac{q_{1,2}^2}
{\mathbf{q}^2(k-q)^2}\frac{1}{q^2+\Pi_{ii}} \nonumber \\
&\approx& -2\frac{\alpha}{N}\int\frac{d^3q}{(2\pi)^3}
\frac{q_{1,2}^2}{\mathbf{q}^2q^2}\frac{1}{\Pi_{ii}} \nonumber \\
&=& -2\frac{\alpha}{N}\int\frac{d^3q}{(2\pi)^3}\frac{q_{1,2}^2}
{\mathbf{q}^2q^2}\frac{8}{\alpha|\mathbf{q}|} \nonumber \\
&=& -\frac{2}{\pi N}\ln\frac{\Lambda_0}{\Lambda_1}.
\end{eqnarray}
Now the total self-energy is
\begin{eqnarray}
\Sigma_{B1} - \Sigma_{B0} &=& (\Sigma_{B1} -
\Sigma_{B0})_{\mathrm{fir}} +
(\Sigma_{B1} - \Sigma_{B0})_{\mathrm{sec}} \nonumber\\
&=&\frac{4}{\pi N}\ln\frac{\Lambda_0}{\Lambda_1}-
\frac{2}{N\pi}\ln\frac{\Lambda_0}{\Lambda_1}\nonumber\\
&=&\frac{2}{N\pi}\ln\frac{\Lambda_0}{\Lambda_1}.\label{B1_B0}
\end{eqnarray}

This result is valid only when $\Lambda_0 > \Lambda_1 > 2\mu$. As
the energy scale decreases below $2\mu$, $\Pi_{ii}$ should be
replaced by its low-energy expression. In this case, the first term
of fermion self-energy becomes
\begin{eqnarray}
(\Sigma_{B1}-\Sigma_{B0})_{\mathrm{fir}} &\approx& \frac{2}{N}
\!\int_{\Lambda_1}^{\Lambda_0}\!\!\frac{d^3q}{(2\pi)^3}
\frac{1}{q^{2}}\frac{1}{q^2+\Pi_{ii}} \nonumber \\
&\approx& \frac{2\alpha}{N}\!\int^{\Lambda_0}_{\Lambda_1}
\!\!\frac{d^3q}{(2\pi)^3}\frac{1}{q^2}\frac{1}{q^2 +
\frac{\mu}{2\pi}\frac{q_0}{|\mathbf{q}|}}.
\end{eqnarray}
In order to simplify calculations, we can divide momenta, frequency,
and chemical potential by $\alpha$ to make all these variables
dimensionless. However, for notational convenience, we still denote
$\Lambda_{0,1}/\alpha$ as $\Lambda_{0,1}$. Now, the above equation
is rewritten as
\begin{eqnarray}
(\Sigma_{B1}-\Sigma_{B0})_{\mathrm{fir}} &=&\frac{2}{N}
\!\int^{\Lambda_0}_{\Lambda_1} \!\!\frac{d^3q}{(2\pi)^3}\frac{1}{q^2}
\frac{1}{q^2 +\frac{\mu}{2\pi}\frac{q_0}{|\mathbf{q}|}}.
\end{eqnarray}
In the low-energy region $q_0 \ll |\mathbf{q}| \ll 2\mu$, this
integral becomes
\begin{eqnarray}
(\Sigma_{B1}-\Sigma_{B0})_{\mathrm{fir}}
&\approx& \frac{2}{N}\int dq_0 \int\frac{d|\mathbf{q}|}
{(2\pi)^2}\frac{1}{|\mathbf{q}|^3+\frac{ \mu}{2\pi}q_0}\nonumber\\
&&\times\theta(\Lambda_0^2 - q_0^2 - \mathbf{q}^2)
\theta(q_0^2 + \mathbf{q}^2 - \Lambda_1^2) \nonumber\\
&\approx& \frac{2}{N}\int dq_0 \int\frac{d|\mathbf{q}|}{(2\pi)^2}
\frac{1}{|\mathbf{q}|^3 + \frac{\mu}{2\pi}q_0}\nonumber\\
&&\times\theta(\Lambda_0^2-\mathbf{q}^2)
\theta(\mathbf{q}^2-\Lambda_1^2) \nonumber\\
&=& \frac{2}{N}\int^{\Lambda_1}_0 \!\!dq_0
\int^{\Lambda_0}_{\Lambda_1}\!\!\frac{d|\mathbf{q}|}{(2\pi)^2}
\frac{1}{|\mathbf{q}|^3 +\frac{ \mu}{2\pi}q_0}. \nonumber\\
\end{eqnarray}
It seems easier to first integrate over $q_0$ and then integrate
over $\mathbf{q}$,
\begin{eqnarray}
(\Sigma_{B1} - \Sigma_{B0})_{\mathrm{fir}} &\approx& \frac{4\pi}{N
\mu} \int^{\Lambda_0}_{\Lambda_1} \frac{d|\mathbf{q}|}{(2\pi)^2} \ln
\left(1+\frac{\frac{\mu}{2\pi}\Lambda_1}
{|\mathbf{q}|^3}\right) \nonumber \\
&\approx& \frac{2}{8\pi^2N}\ln\left(\frac{\Lambda_0} {\Lambda_1}\right) +
\mathrm{others\,\,terms}.\nonumber \\
\end{eqnarray}
Analogously, we have
\begin{eqnarray}
(\Sigma_{B1}-\Sigma_{B0})_{\mathrm{sec}} &=&
-\frac{1}{8\pi^2N}\ln\left(\frac{\Lambda_0}
{\Lambda_1}\right)+\mathrm{others\,\,terms}.\nonumber \\
\end{eqnarray}
We finally obtain the total contribution
\begin{eqnarray}
\Sigma_{B1}-\Sigma_{B0} &=& (\Sigma_{B1} -
\Sigma_{B0})_{\mathrm{fir}} + (\Sigma_{B1} -
\Sigma_{B0})_{\mathrm{sec}} \nonumber \\
&=& \frac{1}{8\pi^2N}\ln \left(\frac{\Lambda_0}
{\Lambda_1}\right).
\end{eqnarray}
Here we keep only the logarithmic term, which survives at the lowest
energy and corresponds to the stable fixed point produced by the
long-ranged transverse gauge interaction.

\begin{figure}
\includegraphics[width=3in]{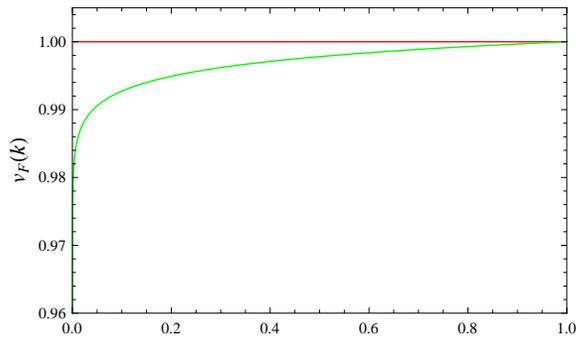}
\vspace{-0.8cm} \caption{Momenta $k$ is rescaled by ultraviolet
cutoff, $k \rightarrow k/\Lambda$. The upper curve represents the
constant velocity at $\mu = 0$. The lower curve shows the
$k$-dependence of renormalized velocity at $\mu \neq
0$.}\label{Fig_v_F}
\end{figure}

Summarizing the above results, we obtain the following RG equation
\begin{eqnarray}
k\frac{dv_F(k)}{dk}=\gamma_{v}v_F(k),
\end{eqnarray}
where
\begin{eqnarray}
\gamma_v&=&\frac{1}{8\pi^2N}.
\end{eqnarray}
This equation has the following solution,
\begin{equation}\label{interaction_v_F}
v_F \propto k^{\gamma_{v}},
\end{equation}
where the constant $\gamma_v$ defines an anomalous dimension of
velocity $v_F$ and the dimensionless momenta $k$ actually
corresponds to $k/\Lambda$ with $\Lambda$ being the UV cutoff.
Obviously, the constant fermion velocity becomes strongly momentum
dependent due to the long-range transverse gauge interaction. The
$k$-dependence of fermion velocity is shown in Fig.~(\ref{Fig_v_F})
at both zero and finite $\mu$. If we define $v_F(\Lambda_0)$ as the
bare velocity, which is taken to be unity in the figure, then the
effective velocity $v_F(k)$ decreases monotonically when $k$ is
lowering. It eventually vanishes as $k \rightarrow 0$.

After this renormalization, the energy spectrum of Dirac fermions
becomes
\begin{equation}\label{eq:dispersion_relation}
\epsilon(k) \propto k^{1+\gamma_{v}}.
\end{equation}
This type of unusual velocity renormalization is a characteristic
feature of Dirac fermion systems, and has been studied extensively
in the contexts of \emph{d}-wave high-temperature superconductor
\cite{Kim97, Huh, WLK} and graphene \cite{Son2007, Sheehy, Vafek}.
The finite anomalous dimension of fermion velocity distinguishes the
system at finite $\mu$ from that at zero $\mu$. An interesting issue
is how this anomalous dimension affects the observable quantities of
Dirac fermions, which will be addressed in the following two
sections.

It is interesting to note that the anomalous dimension $\gamma_v$ is
a universal constant for any given flavor $N$. Although $\gamma_v$
is finite only at finite $\mu$, it does not explicitly depend on
$\mu$. It is the finiteness, rather than the precise value, of $\mu$
that is really important. A finite $\gamma_v$ is generated as long
as $\mu$ becomes finite, no matter how small it is. Therefore, the
ground state changes fundamentally once one moves away from the
neutral Dirac point.

\section{Specific heat of Dirac fermions}\label{sec_specific_heat}

According to Landau FL theory, a weakly interacting fermion system
is in one-to-one correspondence with a non-interacting fermion gas.
There is a sharp Fermi surface in a FL with well-defined
quasiparticles existing in the low energy regime. The properties of
a FL can be manifested in a variety of physical quantities, such as
spectral function, specific heat, DOS, and susceptibility. The
Coulomb interaction does not destroy the stability of FL state in
normal metals, because it is always statically screened by the
collective particle-hole excitations. However, non-FL behavior would
emerge when there is some kind of long-range gauge interaction. We
now would like to examine the corrections of gauge interaction to
several physical quantities of massless Dirac fermions. We will
consider specific heat in this section, and consider DOS and
compressibility in the next section.

In order to calculate specific heat, we first need to calculate the
free energy. In the following, to facilitate calculations we make
the rescaling transformations: $T, k, \mu \rightarrow T/\Lambda,
k/\Lambda, \mu/\Lambda$, where $\Lambda$ is ultraviolet cutoff. When
$\mu = 0$, the specific heat due to gauge interaction has already
been studied in previous works \cite{Kaul, Liu09}, which show that
$C_V \propto T^2$.

The free energy density will be computed using the methods given in
Ref. \cite{Kapusta}. The partition function is
\begin{eqnarray}
Z=\prod_{n,\mathbf{k},\alpha} \int
\mathcal{D}[-i\psi^\dagger_{\alpha,n}(\mathbf{k})]
\mathcal{D}[\psi_{\alpha,n}(\mathbf{k})]e^S,
\end{eqnarray}
where the action is expressed as
\begin{eqnarray}
S &=& \sum_{n,
\mathbf{k}}[-i\psi^\dagger_{\alpha,n}(\mathbf{k})]D_{\alpha,\rho}
[\psi_{\rho,n}(\mathbf{k})],\nonumber\\
D&=&\beta[-(\omega_n-i\mu)-iv_F\gamma^0\gamma\cdot\mathbf{k}].
\end{eqnarray}
Employing the functional integral formula for Grassmann variables,
\begin{eqnarray}
\int \mathcal{D}[\eta^\dagger]\mathcal{D}[\eta]e^{\eta^\dagger D\eta}
=\mathrm{det}D,
\end{eqnarray}
it is easy to obtain
\begin{eqnarray}
\ln Z
&=&\sum_{n,\mathbf{k}} \ln \mathrm{det}D(n,\mathbf{k})\nonumber\\
&=&\sum_{n,\mathbf{k}}\ln\left\{\beta^4\left[(\omega_n-i\mu)^2+
v_F^2(k_1^2+k_2^2)\right]^2\right\}\nonumber\\
&=&\sum_{n,\mathbf{k}}\ln\left\{\beta^4\left[(\omega_n-i\mu)^2+
\epsilon(k)^2\right]^2\right\}\nonumber\\
&=&\sum_{n,\mathbf{k}}
\left\{\ln\left[\beta^2(\omega_n^2+(\epsilon(k)-\mu)^2)\right]\right.\nonumber\\
&&+\left.\ln\left[\beta^2(\omega_n^2+(\epsilon(k)+\mu)^2)\right]\right\}.
\end{eqnarray}
Applying the identities,
\begin{eqnarray}
&&\int^{\beta^2(\omega\pm\mu)^2}_1 \frac{d\theta^2}{\theta^2+(2n+1)^2\pi^2}+
\ln[1+(2n+1)^2\pi^2]\nonumber\\
&&=\ln[(2n+1)^2\pi^2+\beta^2(\omega\pm\mu)^2],\\
&&\sum^\infty_{n=-\infty}\frac{1}{(n-x)(n-y)}
=\frac{\pi[\cot(\pi x)-\cot (\pi y)]}{y-x},
\end{eqnarray}
we are left with
\begin{eqnarray}
\ln Z &=& 2V\int \frac{d^2\mathbf{k}}{(2\pi)^2} \left[\beta
\epsilon(k)+\ln\left(1+e^{-\beta(\epsilon(k)-\mu)}\right)\right.\nonumber\\
&&+\left.\ln\left(1+e^{-\beta(\epsilon(k)+\mu)}\right)\right]
\end{eqnarray}
after some straightforward algebra. Since the free energy density $f
= \frac{F}{V}=-\frac{1}{\beta V}\ln Z$, we have
\begin{equation}\label{eq:f_mu}
f(T,\gamma_{v},\mu) = -2T \int \frac{d^2
\mathbf{k}}{4\pi^2}\ln\left[1+\mathrm{exp} \left(-\frac{\epsilon(k)
\pm \mu}{T}\right)\right],
\end{equation}
where the zero-point energy (the first term in $\ln Z$) has been
discarded. The fermion specific heat can be obtained from free
energy, namely
\begin{eqnarray}
C_V = -T\frac{\partial^2f}{\partial T^2}.
\end{eqnarray}

In the absence of gauge interaction, the fermion velocity takes its
bare value and the anomalous dimension $\gamma_{v} = 0$. Therefore,
the free energy density is simply
\begin{eqnarray}\label{eq:f0f_mu}
f^{0}(\mu) \! &=& \! \!-2T\!\!\int\!\frac{d^2 \mathbf{k}}{4 \pi^2}\!
\left[\ln\!\left(\!1 \!+\!e^{-\frac{k + \mu}{T}} \!\right)\! + \!
\ln \!\left(\!1 \!+ \!e^{-\frac{k- \mu}{T}}\!\right)\! \right].\nonumber\\
\end{eqnarray}
It is easy to integrate over $\mathbf{k}$, and get
\begin{equation}
f^{0}(\mu) = \frac{T^3}{\pi}
\left[\mathrm{Li}_{3}\left(-e^{\frac{\mu}{T}}\right) +
\mathrm{Li}_{3}\left(-e^{-\frac{\mu}{T}}\right) \right],
\end{equation}
where $\mathrm{Li}_{3}(z)$ is polylogarithmic function. The specific
heat has the following form,
\begin{eqnarray}\label{specific_heat_nointeraction_mu_eq0}
C^{0}_V(\mu) &=&\frac{1}{\pi}\left\{\mu^2\ln\left[\left(1+e^{-
\frac{\mu}{T}}\right)\left(1+e^{\frac{\mu}{T}} \right)\right]\right.
\nonumber \\
&&\left.-6T^2\left[\mathrm{Li}_{3}\left( -e^{\frac{\mu}{T}}\right)+
\mathrm{Li}_{3}\left(-e^{-\frac{\mu}{T}}\right)\right]\right.
\nonumber \\
&&\left.-4\mu T\left[\mathrm{Li}_{2}
\left(-e^{-\frac{\mu}{T}}\right) - \mathrm{Li}_{2}
\left(-e^{\frac{\mu}{T}}\right)\right]\right\}.
\end{eqnarray}
It is obvious that the specific heat $C_V^{0} =
\frac{9\zeta(3)}{\pi}T^2\propto T^2$ at $\mu = 0$. At finite $\mu$,
the transverse gauge interaction induces an anomalous dimension for
fermion velocity, which modifies the Dirac fermion energy spectrum.
Now the free energy density becomes
\begin{eqnarray}
f(\mu,\gamma_{v}) \!&=& \!\!-2T \!\int \!\frac{k d k}{2 \pi}\ln \!\left[\!
\left(\!1 \!+\!e^{-\frac{k^{\eta}+\mu}{T}}\!\right)\!\left(\!1\! +\!e^{-\frac{k^{\eta}-\mu}{T}}\!\right)\!\right],\nonumber\\
\end{eqnarray}
where $\eta = 1 + \gamma_{v}$. It is convenient to define $x =
k^\eta$, and write the free energy as
\begin{eqnarray}\label{free_energy_1}
f(\mu,\eta)\! &=&\!\! -\frac{T}{\pi \eta}\!\int\! dx x^{\frac{2-\eta}{\eta}}
\ln\!\left[\!\left(\!1 \!+\! e^{-\frac{x+\mu}{T}}\!\right)\!\left(\!1\!
+\!e^{-\frac{ x -\mu}{T}}\!\right)\!\right].\nonumber\\
\end{eqnarray}
Making derivatives of $f(\mu,\eta)$ with respect to $T$, we obtain
\begin{eqnarray}
C_V \!&=&\!\! \frac{1}{\pi\eta T^2}\!\int\! dx x^{\frac{2-\eta}{\eta}}
\!\left[\frac{(x+\mu)^2 e^{\frac{x+\mu}{T}}}{\left(1 +
e^{\frac{x+\mu}{T}}\right)^2}\!+\! \frac{(x-\mu)^2
e^{\frac{x-\mu}{T}}}{\left(1+e^{\frac{x-\mu}{T}}
\right)^2}\!\right],\nonumber\\\label{specific_heat_interaction_mu_eq0}
\end{eqnarray}
which is complicated and will be evaluated numerically.

\begin{figure}[t]
\centering
       \epsfig{file=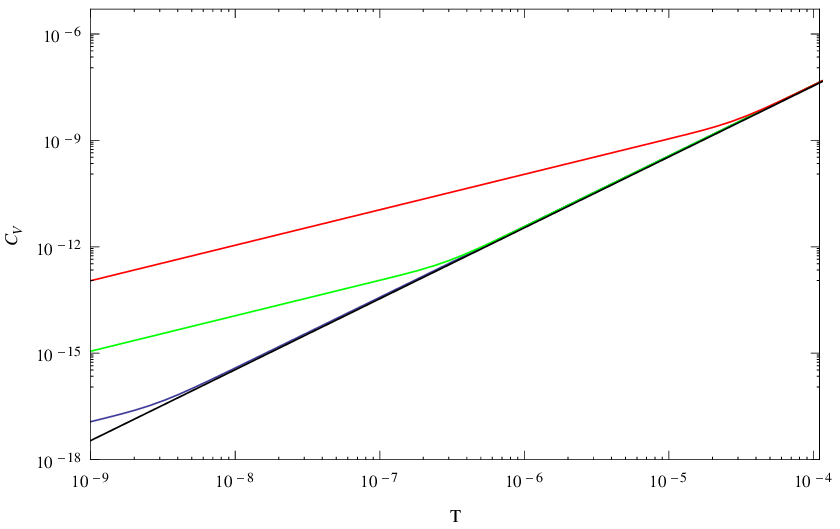,height = 6.18cm,width=8.2cm}
\vspace{-0.35cm} \caption{$\gamma_v=\frac{1}{8\pi^2N}$. The curves
from top to bottom are $C_V(\mu=10^{-4})$, $C_V(\mu=10^{-6})$,
$C_V(\mu=10^{-8})$, and $C_V^{0} =
\frac{9\zeta(3)}{\pi}T^2$, respectively.}\label{fig_C_V_1}
\end{figure}

Note the UV cutoff $\Lambda$ does not qualitatively affect our basic
conclusion, which allows us to set $\Lambda = 1$. The specific heat
$C_V$ explicitly depends on both chemical potential $\mu$ and
temperature $T$. Its $T$-dependence shown in Fig. (\ref{fig_C_V_1})
for $N = 4$. When chemical potential $\mu = 0$, the specific heat
behaves as $C_V \propto T^2$. At finite $\mu$, the specific heat
deviates from the $T^2$ curve. The deviation becomes more
significant for larger $\mu$. At low temperature, the specific heat
can be approximately written as power-law, $C_V \propto T^{\delta}$,
where the exponent $\delta$ is a function of $\mu$. Such
unconventional non-FL behavior arises from the anomalous dimension
of fermion velocity, $\gamma_v$, which is generated by the
long-ranged transverse gauge interaction at finite $\mu$. In
particular, the deviation occurs once $\mu$ becomes finite. This
implies that the ground states of QED$_3$ are very different at zero
and finite $\mu$.


\begin{figure}[t]
\centering
       \epsfig{file=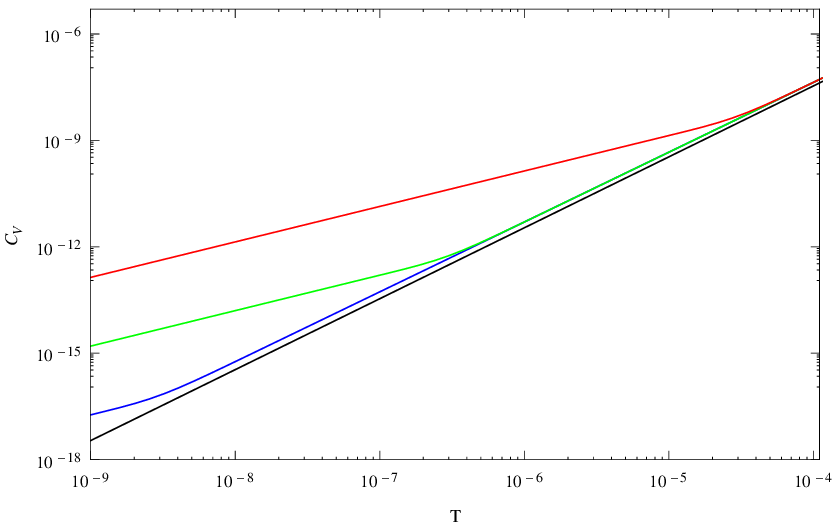,height = 6.18cm,width=8.2cm}
\vspace{-0.35cm} \caption{$\gamma_v\rightarrow5\gamma_v$. The curves
from top to bottom are $C_V(\mu=10^{-4})$, $C_V(\mu=10^{-6})$,
$C_V(\mu=10^{-8})$, and $C_V^{0} =
\frac{9\zeta(3)}{\pi}T^2$, respectively.}\label{fig_C_V_2}
\end{figure}

\begin{figure}[t]
\centering
       \epsfig{file=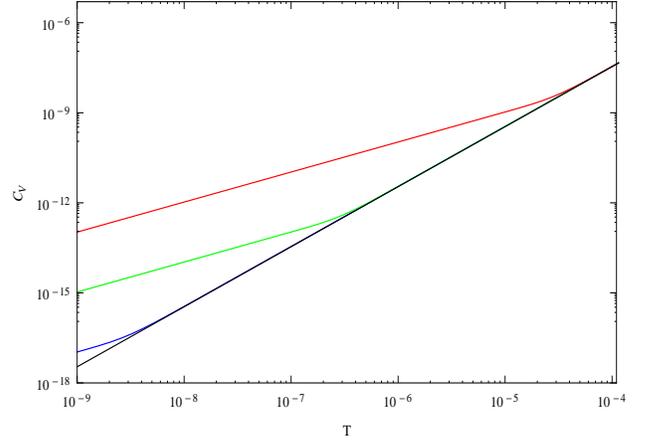,height = 6.18cm,width=8.2cm}
\vspace{-0.35cm} \caption{$\gamma_v\rightarrow\frac{1}{5}\gamma_v$.
The curves from top to bottom are $C_V(\mu=10^{-4})$,
$C_V(\mu=10^{-6})$, $C_V(\mu=10^{-8})$,
and $C_V^{0} =
\frac{9\zeta(3)}{\pi}T^2$, respectively.}\label{fig_C_V_3}
\end{figure}

In many works on non-FL behavior caused by singular interactions,
the specific heat is expressed in a logarithmic function
\cite{Holstein, Reizer, Gan, Kim97, Vafek, Sheehy}, i.e., $C_V
\propto T \ln T$. Note that this logarithmic expression does not
contradict our results. Actually, the power-law specific heat
presented here amounts to a summation of all powers of $\ln T$
\cite{Ipp1, Ipp2, Son2007, Xu}.

In the above calculations, the anomalous dimension $\gamma_v$ plays
an essential role. However, it is obtained by adopting certain
approximations. To examine the reliability of our results, we
suppose $\gamma_{v} = \frac{1}{8\pi^2 N}$ becomes $5\gamma_v$ or
$\gamma_v/5$ after including higher order corrections, and show the
corresponding results in Fig. (\ref{fig_C_V_2}) and Fig.
(\ref{fig_C_V_3}). Apparently, the basic conclusions are independent
of the precise value of anomalous dimension.

\section{Density of states and Compressibility}\label{sec_DOS_kappa}

We now turn to the interaction corrections to DOS and
compressibility. At $\mu=0$, the propagator of massless Dirac
fermion is
\begin{eqnarray}
G(i\omega,\mathbf{k})&=&\frac{1}{i\omega\gamma_0- v_F\gamma\cdot
\mathbf{k}} = \frac{i\omega\gamma_0 + v_F\gamma\cdot\mathbf{k}}
{(i\omega)^2-v_F^2\mathbf{k}^2}.
\end{eqnarray}
After analytical continuation, $i\omega \rightarrow \omega+i\delta$,
we have the following retarded propagator
\begin{eqnarray}
G^{\mathrm{ret}}(\omega,\mathbf{k})
&=&\frac{\omega\gamma_0+v_F\gamma\cdot\mathbf{k}}
{\omega^2-v_F^2\mathbf{k}^2+i\mathrm{sgn}(\omega) \delta}.
\end{eqnarray}
The corresponding spectral function is given by
\begin{eqnarray}
A(\omega,\mathbf{k})
&=&-\frac{1}{\pi}\mathrm{Im}
G^{\mathrm{ret}}(\omega,\mathbf{k})\nonumber\\
&=&\mathrm{sgn}(\omega)(\omega\gamma_0-v_F\gamma
\cdot\mathbf{k})\delta(\omega^2-v_F^2\mathbf{k}^2),
\end{eqnarray}
which then gives rise to the DOS,
\begin{eqnarray}
\rho(\omega)&=& N\int \frac{d^2\mathbf{k}}{(2\pi)^2}
\mathrm{Tr}\{\gamma_0\mathrm{Im}
G^{\mathrm{ret}}(\omega,\mathbf{k})\} \nonumber \\
&=& \frac{N\omega}{v_F^2\pi}.
\end{eqnarray}
Apparently, the DOS vanishes at the Fermi level, $\omega = 0$.

At finite chemical potential $\mu$, the fermion propagator becomes
\begin{eqnarray}
G(i\omega,\mathbf{k})&=&\frac{(i\omega+\mu)
\gamma_0+v_F\gamma\cdot\mathbf{k}}
{(i\omega+\mu)^2-v_F^2\mathbf{k}^2}.
\end{eqnarray}
The DOS can be calculated similarly, with the expression
\begin{eqnarray}\label{No_interaction_Dos_mu_neq_0}
\rho(\omega) &=&N\int \frac{d^2\mathbf{k}}{(2\pi)^2}
\mathrm{Tr}\{\gamma_0
\mathrm{Im}G^{\mathrm{ret}}(\omega,\mathbf{k})\} \nonumber\\
&=& \frac{N(\omega+\mu)}{v_F^2 \pi},
\end{eqnarray}
which approaches a constant proportional to $\mu$ as $\omega
\rightarrow 0$. We are interested in the gauge interaction
corrections to the above expressions. At finite $\mu$, the fermion
velocity $v_F$ becomes $k$-dependent, $v_F'\propto k^{\gamma_{v}}$.
Using this expression, we find that,
\begin{eqnarray}\label{interaction_Dos_mu_neq_0}
\rho(\omega) &=&4N\pi(\omega+\mu)\int \frac{ d k^2}{(2\pi)^2}
\delta((\omega+\mu)^2- v_F^2k^2) \nonumber \\
&=& \frac{ 4N\pi(\omega+\mu)}{\eta v_F^2} \int \frac{d
k'}{(2\pi)^2}k'^{\frac{2 -\eta}{\eta}} \delta((\omega+\mu)^2-k'^2)
\nonumber \\
&=& \frac{ N(\omega+\mu)^{\frac{2 - \eta}{\eta}}}{\eta v_F^2 \pi}.
\end{eqnarray}
The effect of gauge interaction is reflected in the nontrivial
exponent of DOS.

Since the difference between zero and finite chemical potential is
of primary interest in our work, we only consider zero-temperature
compressibility, which is an important quantity describing the
electronic properties of any interacting system. In its original
meaning, the compressibility is defined as $\kappa =
\partial V/\partial F$ \cite{Schwabl}, with $V$ and $F$ being the
volume and compressing force, respectively. However, in practical
many-particle calculations, it is more convenient to define the
compressibility as $\kappa =\partial n/\partial \mu$, where $n$ is
the number of particles per area \cite{Hwang, Sheehy}. The
compressibility of massless Dirac fermions vanishes at zero chemical
potential, $\mu = 0$. This behavior will be changed at finite
density, i.e., $\mu \neq 0$ \cite{Sheehy, Hwang}. In order to obtain
$\kappa$, we only need to calculate the $\mu$-dependence of particle
number $n$.

Using the DOS expressed in Eq. (\ref{No_interaction_Dos_mu_neq_0}),
the particle number in the absence of gauge interaction is
\begin{eqnarray}
n = \int^0_{-\mu}d\omega \rho(\omega) = \int^0_{-\mu}\frac{N(\omega
+ \mu)}{v_F^2 \pi}d\omega = \frac{N\mu^2}{2v_F^2 \pi},\label{n}
\end{eqnarray}
which leads to
\begin{eqnarray}
\kappa=\frac{\partial n}{\partial \mu}
=\frac{N\mu}{v_F^2 \pi}.
\end{eqnarray}
After including the gauge interaction, the DOS is given by Eq.
(\ref{interaction_Dos_mu_neq_0}). Now the particle number becomes
\begin{eqnarray}
n=\int^0_{-\mu}d\omega\rho(\omega)
=\int^0_{-\mu} \frac{N(\omega+\mu)^{\frac{2-\eta}{\eta}}}{\eta v_F^2\pi}d\omega
=\frac{N\mu^{\frac{2}{\eta}}}{2v_F^2\pi},
\end{eqnarray}
which then yields
\begin{eqnarray}
\kappa=\frac{N\mu^{\frac{2}{\eta}-1}}{\eta v_F^2 \pi} .
\end{eqnarray}
Once again, the effect of gauge interaction is reflected in the
exponent.

\section{Summary}\label{sec_summary}

In summary, we have studied the effects of a finite chemical
potential on the behavior of Dirac fermions in QED$_3$. At zero
chemical potential, there is no fermion velocity renormalization. At
finite chemical potential, the longitudinal gauge interaction
becomes short-ranged, but the transverse gauge interaction remains
long-ranged and leads to singular velocity renormalization. An
explicit calculation shows that a finite anomalous dimension of
velocity is generated and gives rise to unconventional properties in
some physical quantities, including specific heat, DOS, and
compressibility. Therefore, the massless Dirac fermions behave quite
differently at finite and zero chemical potential. This difference,
together with the difference in fermion damping rate \cite{WangLiu1,
WangLiu2}, indicates that the ground state of QED$_3$ is
fundamentally changed once the chemical potential becomes finite.

\section*{Acknowledgments}

J.W. is grateful to W. Li and J.-R. Wang for their help. G.Z.L.
acknowledges financial support by the National Natural Science
Foundation of China under Grant No. 11074234 and the Visitors
Program of MPIPKS at Dresden.

\end{document}